\documentclass[journal, one column]{IEEEtran}
\usepackage{amsmath, amsthm, amsfonts}
\usepackage[english]{babel}
\usepackage[dvips]{graphicx}
\usepackage{graphicx}
\usepackage{tabularx}
\usepackage{enumerate}
\usepackage{mathtools}

\usepackage{setspace}
\begin{document}
\title{Convergence Analysis of Proportionate-type Least Mean Square Algorithms}


\author{{Vinay Chakravarthi Gogineni$^1$, Subrahmanyam Mula$^2$}\\
Department of Electronics and Electrical Communication Engineering\\
Indian Institute of Technology, Kharagpur, INDIA\\
E.Mail : $^1\;$vinaychakravarthi@ece.iitkgp.ernet.in, $^2\;$svmula@iitkgp.ac.in}

\maketitle
\thispagestyle{empty}

\begin{abstract}
In this paper, we present the convergence analysis of proportionate-type least mean square (Pt-LMS) algorithm that identifies the sparse system effectively and more suitable for real time VLSI applications. Both first and second order convergence analysis of Pt-LMS algorithm is studied. Optimum convergence behavior of Pt-LMS algorithm is studied from the second order convergence analysis provided in this paper. Simulation results were conducted to verify the analytical results.
\end{abstract}
\section{Introduction}
Usually, many real-life systems exhibit sparse representation
i.e., their system impulse response is characterized by small
number of non zero taps in the presence of large number of
inactive taps. Sparse systems are encountered in many important
practical applications such as network and acoustic echo
cancelers $\cite{1}$-$\cite{2}$, HDTV channels $\cite{3}$,
wireless multipath channels $\cite{4}$, underwater acoustic
communications $\cite{5}$. The conventional
system identification algorithms such as LMS and NLMS
are sparsity agnostic i.e., they are unaware of
underlying sparsity of the system impulse response.  Recent
studies have shown that the \emph{a priori} knowledge about the
system sparsity, if utilized properly by the identification
algorithm, can result in substantial improvement in its estimation
performance. This resulted in a flurry of research activities in
the last decade or so towards developing sparsity aware adaptive
filter algorithms, notable amongst them being the Proportionate
Normalized LMS (PNLMS) algorithm $\cite{6}$ and its variants
$\cite{7}$-$\cite{9}$. Unlike the NLMS, the weighted Euclidean norm of the input vector presented in Proportionate-type NLMS (Pt-NLMS) can not be computed recursively due to the presence of gain matrix $\textbf{G}(n)$, which varies at each time instance $n$. Computation of this weighted Euclidean norm of the input vector requires requires $2N$ multiplications and $N-1$ additions in each iteration that limits the throughput for real-time applications. In this paper, we present the performance analysis of Proportionate-type LMS (Pt-LMS) algorithm that is more suitable for real time VLSI applications.
\section{Proportionate-type LMS Algorithm}
We consider the problem of identifying an unknown system (supposed to
be sparse), modeled by the $L$ tap coefficient vector $\textbf{w}_{opt}$
which takes a signal $u(n)$ with variance $\sigma_{u}^{2}$
as the input and produces the observable output
$d(n)=\textbf{u}^{T}(n) \textbf{w}_{opt} + v(n)$, where
$\textbf{u}(n)=[ {u}(n), {u}(n-1), . . . , {u}(n-L+1)]^{T}$ is the
input data vector at time $n$, and $v(n)$ is the observation
noise with zero mean and variance $\sigma_{v}^{2}$ which is assumed to be white and independent of
$u(m)$ for all $n,\;m$. The Pt-LMS algorithm iteratively updates
the filter coefficient vector $\textbf{w}=[w_{0}, w_{1}, . . .,
w_{L-1}]^{T}$ as,
\begin{equation}\label{eq1.1}
\textbf{w}(n+1)=\textbf{w}(n)+ \mu \hspace{0.2em} \textbf{G}(n) \hspace{0.2em} \textbf{u}(n) \hspace{0.2em} e(n)
\end{equation}
where  $\mu$ is the step size, $\textbf{G}(n)$ is a diagonal gain
matrix that distributes the adaptation energy unevenly over the
filter taps by modifying the step size of each tap, and
$e(n)=d(n)-\textbf{u}^{T}(n)\textbf{w}(n)$ is the filter output
error.
\par

The gain matrix $\textbf{G}(n)$ is evaluated as,
\begin{equation}\label{eq1.2}
 \textbf{G}(n)=diag(g_{0}(n),g_{1}(n),...g_{L-1}(n))
\end{equation}
where,
\begin{equation}\label{eq1.3}
g_{l}(n)=\frac{\gamma_{l}(n)}{\frac{1}{L} \sum\limits_{l=0}^{L-1}\gamma_{l}(n)}, \mspace{8mu} 0\leq l \leq (L-1)
\end{equation}
with,
\begin{equation}\label{eq1.4}
\gamma_{ l}(n)=max[\rho\mspace{4mu} \gamma_{min}(n), \mathrm{F}[|w_{l}(n)|]]
\end{equation}
\begin{equation}\label{eq1.5}
\gamma_{min}(n)=max(\delta, \mathrm{F}[|w_{0}(n)|],...,\mathrm{F}[|w_{L-1}(n)|]
\end{equation}
where $\rho$ is a very small, positive constant which, together
with $\gamma_{min}(n)$, ensures that $\gamma_{l}(n)$ and thus
$g_{l}(n)$ do not turn out to be zero for the inactive taps and
thus the corresponding updation does not stall. The parameter
$\delta$ is again a small positive constant employed to avoid
stalling of the weight updation at the start of the iterations
when the tap weight iterates are initialized to zero. The function
$\mathrm{F}[|w_{l}(n)|]$ is chosen differently for different
Pt-LMS algorithms, as described in the table below. From (1), it
is easily seen that $\mu g_l(n)$ provides the effective step size
for the $l$-th tap which, through the function
$\mathrm{F}[|w_{l}(n)|]$, is monotonically related to $|w_l(n)|$.
\begin{table}[h!]
\centering
\caption {The function $\mathrm{F}[|w_{l}(n)|]$ for a few popular
Pt-LMS algorithms}
    \begin{tabular}{ | l | p{4.5cm} |}
    \hline \\[-1.0em]
    Algorithm & $\mathrm{F}[|w_{l}(n)|]$ or $g_{l}(n)$ \\ [ 2ex]\hline \\[-1.0em]
    1. Standard LMS & $\mathrm{F}[|w_{l}(n)|] =1$ \\ [ 2ex]\hline \\[-1.0em]
    2. PLMS & $\mathrm{F}[|w_{l}(n)|] = |w_{l}(n)|$ \\ [ 2ex] \hline \\[-1.0em]
    3. IPLMS & $g_{l}(n) = \frac{1-\alpha}{2N} +  \frac{1+\alpha}{2}\frac{|w_{l}(n)|}{\|w_{l}(n)\|_{1}+\delta_{I}}$\hspace{1em} $\alpha \in [-1, 1]$ \\ [ 2ex] \hline \\[-1.0em]
    3. $\mu$-law PLMS & $\mathrm{F}[|w_{l}(n)|]=ln(\frac{1+( \epsilon |w_{l}(n)|)}{1+\epsilon})$; \hspace{1em} $\epsilon$ is a positive constant \\ \hline

    \end{tabular}
\end{table}
\section{Performance Analysis of Proportionate-type Least Mean Square Algorithm}
In this section, we examine the convergence behavior of the
proposed proportionate-type least mean square algorithm.
\subsection{Mean Convergence Analysis of Pt-LMS Algorithm}
By denoting $\widetilde{\textbf{w}}(n)= \textbf{w}_{opt}-\textbf{w}(n)$, from \eqref{eq1.1} the recursion for the weight error vector of the Pt-LMS algorithm can be written as follows:
\begin{equation}\label{eq2.1}
\begin{split}
\widetilde{\textbf{w}}(n+1)&=\Big[\textbf{I}_{L}-\mu \hspace{0.2em} \textbf{G}(n) \hspace{0.2em} \textbf{u}(n) \hspace{0.2em} \textbf{u}^{T}(n) \Big] \widetilde{\textbf{w}}(n)
 -\mu \hspace{0.2em} \textbf{G}(n) \hspace{0.2em} \textbf{u}(n) \hspace{0.2em} v(n)
\end{split}
\end{equation}
The equation \eqref{eq2.1} forms the basis for the performance analysis of the Pt-LMS algorithm. Using the statistical independence between
$\textbf{w}(n)$ and $\textbf{u}(n)$ (i.e., ``independence assumption"), and recalling that $v(n)$ is zero-mean i. i. d random variable which is independent of $\textbf{u}(n)$ and thus of $\widetilde{\textbf{w}}(n)$, one can write
\begin{equation}\label{eq2.2}
\begin{split}
E[\widetilde{\textbf{w}}(n+1)]&=\Big[\textbf{I}_{L}-\mu \hspace{0.2em} E\big[\textbf{G}(n) \hspace{0.2em} \textbf{u}(n) \hspace{0.2em} \textbf{u}^{T}(n) \big]\Big] E[\widetilde{\textbf{w}}(n)]
\end{split}
\end{equation}
When compared to $\textbf{w}(n)$ as $\textbf{G}(n)$ changes slowly with time (nearly convergence), we can assume $\textbf{G}(n)$ is independent of $\textbf{u}(n)$. Therefore, the above equation can be rewritten as,
\begin{equation}\label{eq2.3}
\begin{split}
E[\widetilde{\textbf{w}}(n+1)]&=\Big[\textbf{I}_{L}-\mu \hspace{0.2em} \overline{\textbf{G}} \hspace{0.2em} \textbf{R}\Big] E[\widetilde{\textbf{w}}(n)]
\end{split}
\end{equation}
where $E[\textbf{G}(n)=\overline{\textbf{G}}$. From the above result, the convergence of Pt-LMS family is guaranteed only if and only if
\begin{equation}\label{eq2.4}
\begin{split}
| \lambda_{max}\big( \textbf{I}_{L}-\mu \hspace{0.2em} \overline{\textbf{G}} \hspace{0.2em} \textbf{R} \big) |< 1
\end{split}
\end{equation}
Therefore, a sufficient condition for $\eqref{eq2.4}$ to hold is
\begin{equation}\label{eq2.5}
\begin{split}
0 < \mu < \frac{2}{\lambda_{max}\big(\overline{\textbf{G}} \hspace{0.2em} \textbf{R} \big)}
\end{split}
\end{equation}
From matrix norm inequalities, finally the condition on $\mu$ is
\begin{equation}\label{eq2.6}
\begin{split}
0 < \mu < \frac{2}{\overline{g}_{max} \hspace{0.2em} \lambda_{max}\big( \hspace{0.2em} \textbf{R} \big)}
\end{split}
\end{equation}
For white regressor data for which $\textbf{R}= \sigma^{2}_{u} \textbf{I}$, from \cite{10} we have $Tr(\overline{\textbf{G}} \textbf{R}) =1$. Therefore, for white input signal case, a sufficient condition for $\eqref{eq2.4}$ to hold is  $0 < \mu < 2$. 
\subsection{Mean-Square Error Behavior Analysis}
Using the statistical independence between $\textbf{w}(n)$ and $\textbf{u}(n)$ (i.e., ``independence assumption"), and recalling that $v(n)$ is of zero-mean and also independent of $\textbf{u}(n)$ and thus of $\widetilde{\textbf{w}}(n)$, from $\eqref{eq2.1}$, using energy conservation approach \cite{11}, the mean square of the weight error vector $\widetilde{\textbf{w}}(n)$, weighted by any positive semi-definite matrix $\boldsymbol{\Sigma}$ that we are free to choose, satisfies the following relation :
\begin{equation}\label{eq2.7}
\begin{split}
E\|\widetilde{\textbf{w}}(n+1)\|_{\boldsymbol{\Sigma}}^{2}&= E\|\widetilde{\textbf{w}}(n)\|_{E \boldsymbol{\Sigma}^{'}}^{2}
+ \mu^{2} \hspace{0.2em} E[v^{2}] \hspace{0.2em} E\big[\textbf{u}^{T}(n) \textbf{G}(n) \Sigma \textbf{G}(n) \textbf{u}(n) \big]
\end{split}
\end{equation}
where
\begin{equation}\label{eq2.8}
\begin{split}
\boldsymbol{\Sigma}^{'}&= \boldsymbol{\Sigma} - \mu \hspace{0.2em} \textbf{u}(n) \textbf{u}^{T}(n) \hspace{0.2em} \textbf{G}(n) \hspace{0.2em} \boldsymbol{\Sigma} - \mu \hspace{0.2em} \boldsymbol{\Sigma} \hspace{0.2em} \textbf{G}(n) \hspace{0.2em} \textbf{u}(n) \textbf{u}^{T}(n) + \mu^{2} \hspace{0.2em} \textbf{u}(n) \textbf{u}^{T}(n) \hspace{0.2em} \textbf{G}(n) \hspace{0.2em} \boldsymbol{\Sigma}  \hspace{0.3em} \textbf{G}(n) \hspace{0.2em} \textbf{u}(n) \textbf{u}^{T}(n)\\
\end{split}
\end{equation}
The relations presented in $\eqref{eq2.7}$ and $\eqref{eq2.8}$ are useful to derive the condition for mean square stability and expressions for MSE and MSD. To extract the matrix $\boldsymbol{\Sigma}$ from the expectation terms, a weighted variance relation is introduced by using $L^{2} \times 1$ column vectors:
\begin{equation}\label{eq2.9}
\begin{split}
\boldsymbol{\sigma}= \text{vec}\{\boldsymbol{\Sigma}\} \hspace{2em} \text{and} \hspace{2em} \boldsymbol{\sigma^{'}}= \text{vec}\{E\boldsymbol{\Sigma}^{'}\}
\end{split}
\end{equation}
where $\text{vec}\{\cdot\}$ denotes the vector operator. In addition, $\text{vec}\{\cdot\}$ is also used to recover the original matrix $\Sigma$ from $\boldsymbol{\sigma}$. One property of the $\text{vec}\{\cdot\}$ operator when working with the Kronecker product \cite{12} is used in this work, namely,
\begin{equation}\label{eq2.10}
\begin{split}
\text{vec}\{\textbf{Q} \boldsymbol{\Sigma} \textbf{P}\}= ( \textbf{P}^{T} \otimes \textbf{Q} ) \hspace{0.2em} \boldsymbol{\sigma}
\end{split}
\end{equation}
where $\textbf{P} \otimes \textbf{Q}$ denotes the Kronecker product of two matrices.
\par
Using $\eqref{eq2.10}$ to $\eqref{eq2.8}$ after vectorization, a linear relation between the corresponding vectors $\{\boldsymbol{\sigma},\boldsymbol{\sigma}^{'}\}$ is formulated as follows:
\begin{equation}\label{eq2.11}
\begin{split}
\boldsymbol{\sigma}^{'}= \textbf{F} \boldsymbol{\sigma}
\end{split}
\end{equation}
where the coefficient matrix $\textbf{F}$ is $L^{2} \times L^{2}$ and defined as
\begin{equation}\label{eq2.12}
\begin{split}
\textbf{F} = \textbf{I} - \mu \hspace{0.2em} \big(\textbf{I} \otimes \textbf{R} \big) \hspace{0.2em} \big(\textbf{I} \otimes \overline{\textbf{G}} \big) -\mu \hspace{0.2em} \big( \textbf{R} \otimes \textbf{I} \big) \hspace{0.2em} \big( \overline{\textbf{G}} \otimes \textbf{I} \big) + \mu^{2} \hspace{0.2em} \boldsymbol{\Pi} \hspace{0.2em} E\big(\textbf{G} \otimes \textbf{G} \big)
\end{split}
\end{equation}
with $\boldsymbol{\Pi}= E\Big[\big(\textbf{u}(n) \textbf{u}^{T}(n) \big) \otimes \big(\textbf{u}(n) \textbf{u}^{T}(n)\big) \Big]$.
\par
The term $E[v^{2}] \hspace{0.2em} E\big[\textbf{u}^{T}(n) \textbf{G}(n) \Sigma \textbf{G}(n) \textbf{u}(n) \big]$ can be written as
\begin{equation}\label{eq2.13}
\begin{split}
E[v^{2}] \hspace{0.2em} E\big[\textbf{u}^{T}(n) \textbf{G}(n) \boldsymbol{\Sigma} \textbf{G}(n) \textbf{u}(n) \big]&=\sigma^{2}_{v} \hspace{0.2em} Tr\bigg(E\Big[ \textbf{G}(n) \hspace{0.2em} \textbf{u}(n)\textbf{u}^{T}(n) \hspace{0.2em} \textbf{G}(n) \Big] \boldsymbol{\Sigma}  \bigg)\\
&= \sigma^{2}_{v} \hspace{0.2em} \boldsymbol{\gamma}^{T} \hspace{0.2em} \boldsymbol{\sigma}
\end{split}
\end{equation}
where
\begin{equation}\label{eq2.14}
\begin{split}
\boldsymbol{\gamma}&= \text{vec}\Big\{E\big[ \textbf{G}(n) \hspace{0.2em} \textbf{u}(n)\textbf{u}^{T}(n) \hspace{0.2em} \textbf{G}(n) \big]\Big\}\\
&=E \big(\textbf{G} \otimes \textbf{G} \big) \hspace{0.2em}\boldsymbol{\gamma}_{R}
\end{split}
\end{equation}
with $\boldsymbol{\gamma}_{R}= \text{vec}\{R\}$. Using these results the recursion presented in $\eqref{eq2.7}$ can be rewritten as
\begin{equation}\label{eq2.15}
\begin{split}
E\|\widetilde{\textbf{w}}(n+1)\|_{\boldsymbol{\sigma}}^{2}&= E\|\widetilde{\textbf{w}}(n)\|_{\textbf{F} \hspace{0.2em} \boldsymbol{\sigma}}^{2}
+ \mu^{2} \hspace{0.2em} \sigma^{2}_{v} \hspace{0.2em} \boldsymbol{\gamma}^{T} \hspace{0.2em} \boldsymbol{\sigma}
\end{split}
\end{equation}
The Pt-LMS algorithms are mean square stable if, and only if, the matrix $\textbf{F}$ is stable. Iterating the above recursion starting from $n=0$, we get
\begin{equation}\label{eq2.16}
\begin{split}
E\|\widetilde{\textbf{w}}(n+1)\|_{\boldsymbol{\sigma}}^{2}&= E\|\widetilde{\textbf{w}}(0)\|^{2}_{\textbf{F}^{n+1} \hspace{0.2em} \boldsymbol{\sigma}}
+ \mu^{2} \hspace{0.2em} \sigma^{2}_{v} \hspace{0.2em} \boldsymbol{\gamma}^{T} \hspace{0.1em} \sum\limits_{i=0}^{n} \textbf{F}^{i}\boldsymbol{\sigma}
\end{split}
\end{equation}
Therefore, by selecting $\boldsymbol{\Sigma}=\textbf{I}$, we can relate $E\|\widetilde{\textbf{w}}(n+1)\|_{\boldsymbol{\sigma}}^{2}$ and $E\|\widetilde{\textbf{w}}(n)\|_{\boldsymbol{\sigma}}^{2}$ as follows:
\begin{equation}\label{eq2.16}
\begin{split}
E\|\widetilde{\textbf{w}}(n+1)\|_{\boldsymbol{\sigma}}^{2}&= E\|\widetilde{\textbf{w}}(n)\|_{\boldsymbol{\sigma}}^{2}- E\|\widetilde{\textbf{w}}(0)\|^{2}_{[\textbf{I}-\textbf{F}]\textbf{F}^{n} \hspace{0.2em} \boldsymbol{\sigma}}
+ \mu^{2} \hspace{0.2em} \sigma^{2}_{v} \hspace{0.2em} \boldsymbol{\gamma}^{T} \hspace{0.1em}  \textbf{F}^{n}\boldsymbol{\sigma}
\end{split}
\end{equation}
The weighted variance relation is useful to characterize the transient behavior of the Pt-LMS family. It is also useful to examine the steady-state MSD, which is given as follows:
\begin{equation}\label{eq2.18}
\begin{split}
\lim\limits_{n \to \infty}E\|\widetilde{\textbf{w}}(n)\|_{\big(I_{L^{2}}- \textbf{F} \big) \boldsymbol{\sigma}}^{2}&= \mu^{2} \hspace{0.2em} \sigma^{2}_{v} \hspace{0.2em} \boldsymbol{\gamma}^{T} \boldsymbol{\sigma}
\end{split}
\end{equation}
By selecting $\boldsymbol{\Sigma}=\textbf{I}_{L}$, the steady-state MSD is given as
\begin{equation}\label{eq2.19}
\begin{split}
\lim\limits_{n \to \infty}E\|\widetilde{\textbf{w}}(n)\|^{2}&= \mu^{2} \hspace{0.2em} \sigma^{2}_{v} \hspace{0.2em} \boldsymbol{\gamma}^{T} \big(\textbf{I}_{L^{2}}- \textbf{F} \big)^{-1} \hspace{0.2em} \text{vec}\{ \textbf{I} \}
\end{split}
\end{equation}
Let $\textbf{C}=\big[\big(\textbf{I} \otimes \textbf{R} \big) \hspace{0.2em} \big(\textbf{I} \otimes \overline{\textbf{G}} \big)\big] + \big[ \big( \textbf{R} \otimes \textbf{I} \big) \hspace{0.2em} \big( \overline{\textbf{G}} \otimes \textbf{I} \big) \big]$ and $\textbf{D}= \boldsymbol{\Pi} \hspace{0.2em} E\big(\textbf{G} \otimes \textbf{G} \big)$ so that $\textbf{F}= \textbf{I}_{L^{2}}- \mu \hspace{0.2em} \textbf{C} +\mu^{2} \hspace{0.2em} \textbf{D}$.
\par
From \cite{10},the convergence in the mean square sense of Pt-LMS family is guaranteed for any $\mu$ in the range
\begin{equation}\label{eq2.17}
\begin{split}
0 < \mu < \text{min}\bigg\{ \frac{1}{\lambda_{max}(\textbf{C}^{-1} \hspace{0.2em}\textbf{D})}, \frac{1}{\text{max}(\lambda(\textbf{H}))} \bigg\}
\end{split}
\end{equation}
where $H=\left[ \begin{array}{c} \frac{1}{2} \textbf{C}  \hspace{1em} -\frac{1}{2} \textbf{D}\\ \textbf{I} \hspace{2.5em} 0 \end{array} \right]$.

\section{Simulation Studies and Discussion}
Here the Simulation results are presented for system identification example. First, the proposed algorithm has been simulated for identifying the system ($\textbf{w}_{opt}$) of length $L=512$ having $64$ active taps with the remaining coefficients being inactive as shown in Fig. 1. 
\begin{figure}[h!]
\centering
\includegraphics [height=30mm,width=85mm]{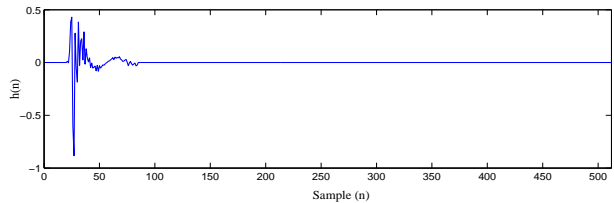}
\caption{Sparse system impuse response}
\label{the-label-for-cross-referencing}
\end{figure}
\par
Simulations were performed using zero mean, Gaussian white noise with unit variance ($\sigma^{2}_{u}=1$ ). The observation noise $v(n)$ was taken to be zero-mean Gaussian white noise with variance $\sigma^{2}_{v}=0.01$. The performance of the proposed Pt-LMS algorithm was compared with the existing PNLMS and LMS algorithms by plotting the respective learning curves (i.e., normalized MSD in dB vs no. of iterations) which are shown in Fig. 2. The simulation results shown in Fig. $2$ are obtained by plotting the normalized MSD against the iteration index $n$, by averaging over $200$ experiments.
\par
\begin{figure}[h!]
\centering
\includegraphics [height=80mm,width=85mm]{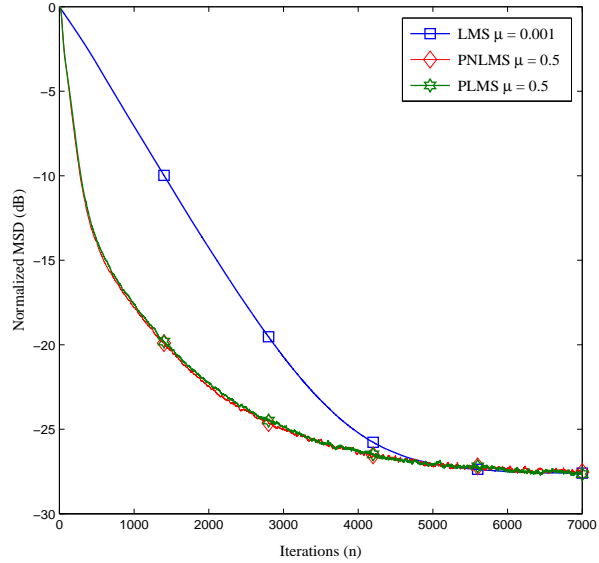}
\caption{Leaning curves of Pt-LMS}
\label{the-label-for-cross-referencing}
\end{figure}
\par
Secondly, for theoretical performance comparison purpose, we considered a sparse system of length $32$ having $2$ active taps with the remaining coefficients being inactive. The input was taken to be zero mean, Gaussian white noise with unit variance ($\sigma^{2}_{u}=1$ ). The observation noise $v(n)$ was taken to be zero-mean Gaussian white noise with variance $\sigma^{2}_{v}=0.01$. Theoretical and simulation results were compared by plotting the steady-state normalized MSD in dB vs step size value ($\mu$) which are shown in Fig. 3. From Fig. 3, we can see the analytical results are coinciding with simulation results.
\begin{figure}[h!]
\centering
\includegraphics [height=50mm,width=90mm]{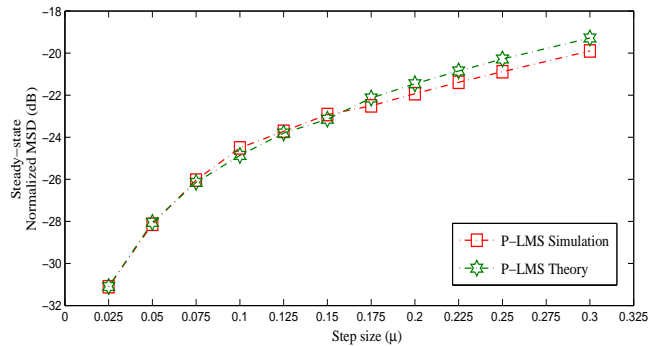}
\caption{Steady-state MSD comparison}
\label{the-label-for-cross-referencing}
\end{figure}
\section{Conclusions}
We presented the performance analysis of Proportionate-type LMS (Pt-LMS) algorithm that is more suitable for real time VLSI applications. The convergence analysis of Pt-LMS algorithm is studied in mean and mean-square sense.

\end{document}